\documentclass[12pt,a4paper]{article}

\usepackage[T1]{fontenc}
\usepackage[utf8]{inputenc}
\usepackage[nosort]{cite}
\usepackage{relsize}
\usepackage{color}
\usepackage[bulletsep]{collref}
\usepackage{array}
\usepackage{graphicx}
\usepackage{bbm}
\usepackage{amsmath}
\usepackage{amssymb}
\usepackage{subfig}
\usepackage{verbatim}
\usepackage{multirow}
\usepackage{tikz}

\usetikzlibrary{shapes.geometric}
\usetikzlibrary{calc,patterns,angles,quotes}
\usetikzlibrary{decorations.pathreplacing}
\usepackage{amsthm}
\usepackage{fancyhdr}
\usepackage{wrapfig}
\usepackage{hyperref}
\usepackage{dsfont}
\usepackage[extdef]{delimset} 
\usepackage{rotating}
 \usepackage{slashed}

\makeatletter \@addtoreset{equation}{section} \makeatother

\makeatletter
\let\old@startsection=\@startsection
\let\oldl@section=\l@section
\renewcommand{\@startsection}[6]{\old@startsection{#1}{#2}{#3}{#4}{#5}{#6\mathversion{bold}}}
\renewcommand{\l@section}[2]{\oldl@section{\mathversion{bold}#1}{#2}}
\makeatother

\makeatletter
\let\old@makecaption=\@makecaption
\def\@makecaption{\small\old@makecaption}
\makeatother

\let\oldPhi=\Phi
\let\oldPsi=\Psi
\let\oldGamma=\Gamma
\let\oldDelta=\Delta
\let\oldSigma=\Sigma
\let\oldtheta=\theta
\let\oldPi=\Pi
\let\oldUpsilon=\Upsilon
\renewcommand{\Phi}{\mathnormal{\oldPhi}}
\renewcommand{\Psi}{\mathnormal{\oldPsi}}
\renewcommand{\Gamma}{\mathnormal{\oldGamma}}
\renewcommand{\Sigma}{\mathnormal{\oldSigma}}
\renewcommand{\Delta}{\mathnormal{\oldDelta}}
\renewcommand{\theta}{\mathnormal{\oldtheta}}
\renewcommand{\Pi}{\mathnormal{\oldPi}}
\renewcommand{\Upsilon}{\mathnormal{\oldUpsilon}}

\newcommand{\superN}{\mathcal{N}}

\newcommand{\sign}{\mathop{\mathrm{sign}}}

\newcommand{\gen}[1]{\mathrm{#1}}
\newcommand{\levo}[1]{ \gen{\widehat #1}}
\newcommand{\Eval}{s} 

\newcommand{\dd}{\mathrm{d}}

\def\sp{\hspace{0.05cm}}

\def\spac{\hspace{1cm}}

\makeatletter
\newlength{\apb@width}
\newcommand{\autoparbox}[2][c]{\settowidth{\apb@width}{#2}\parbox[#1]{\apb@width}{#2}}
\newcommand{\includegraphicsbox}[2][]{\autoparbox{\includegraphics[#1]{#2}}}
\makeatother

\ifx\genfrac\sdflkaj

\else

\fi
\newcommand{\sfrac}[2]{{\textstyle\frac{#1}{#2}}}
\newcommand{\half}{\sfrac{1}{2}}

\newcommand{\signblockbr}[3]{
\begingroup 
\renewcommand*{\arraystretch}{.3}
\begin{bmatrix}
\scriptscriptstyle#1\\
\scriptscriptstyle#2\\
\scriptscriptstyle#3
\end{bmatrix}
\endgroup
}
\newcommand{\signblock}[3]{
\begingroup 
\renewcommand*{\arraystretch}{.3}
\begin{matrix}
\scriptscriptstyle#1\\
\scriptscriptstyle#2\\
\scriptscriptstyle#3
\end{matrix}
\endgroup
}
\newcommand{\signblockbiggerbr}[3]{
\begingroup 
\renewcommand*{\arraystretch}{.3}
\begin{bmatrix}
#1\\
#2\\
#3
\end{bmatrix}
\endgroup
}
\newcommand{\signblocktextbiggerbr}[3]{
\begingroup 
\renewcommand*{\arraystretch}{1}
\begin{bmatrix}
#1\\
#2\\
#3
\end{bmatrix}
\endgroup
}
\newcommand{\signblocktextbiggerbralt}[6]{
\begingroup 
\renewcommand*{\arraystretch}{1}
\begin{bmatrix}
#1&#2\\
#3&#4\\
#5&#6
\end{bmatrix}
\endgroup
}

\makeatletter
\def\mr@ignsp#1 {\ifx\:#1\@empty\else #1\expandafter\mr@ignsp\fi}%
\newcommand{\multiref}[1]{\begingroup
\xdef\mr@no@sparg{\expandafter\mr@ignsp#1 \: }%
\def\mr@comma{}%
\@for\mr@refs:=\mr@no@sparg\do{\mr@comma\def\mr@comma{,}\ref{\mr@refs}}%
\endgroup}
\makeatother

\newcommand{\hypref}[2]{\ifx\href\asklfhas #2\else\href{#1}{#2}\fi}
\newcommand{\Secref}[1]{Section~\multiref{#1}}

\newcommand{\Appref}[1]{Appendix~\multiref{#1}}

\newcommand{\Tabref}[1]{Table~\multiref{#1}}

\newcommand{\Figref}[1]{Figure~\multiref{#1}}

\renewcommand{\eqref}[1]{(\multiref{#1})}

\ifx\href\asklfhas\newcommand{\href}[2]{#2}\fi

\newcommand{\be}{\begin{eqnarray}}
\newcommand{\ee}{\end{eqnarray}}

\begin{document}

\thispagestyle{empty}


\begin{flushright}\footnotesize
\texttt{HU-Mathematik-2020-07}\\
\texttt{HU-EP-20/38}\\
\texttt{SAGEX-20-27-E}
\end{flushright}

\begin{center}%
{\LARGE\textbf{\mathversion{bold}%
Minkowski Box from Yangian Bootstrap
}\par}
\medskip

\vspace{1.2cm}

 \textsc{Luke Corcoran, Florian Loebbert,\\ Julian Miczajka, Matthias Staudacher} \vspace{8mm} \\
\textit{%
Institut f\"{u}r Physik, Humboldt-Universit\"{a}t zu Berlin, \\
Zum Gro{\ss}en Windkanal 6, 12489 Berlin, Germany
} \\

\texttt{\\ \{corcoran,loebbert,miczajka,staudacher\}@physik.hu-berlin.de}
\par\vspace{20mm}

\textbf{Abstract}
 \vspace{5mm}

\begin{minipage}{12cm}
We extend the recently developed Yangian bootstrap for Feynman integrals to Minkowski space, focusing on the case of the one-loop box integral. The space of Yangian invariants is spanned by the Bloch--Wigner function and its discontinuities. Using only input from symmetries, we constrain the functional form of the box integral in all 64 kinematic regions up to twelve (out of a priori 256) undetermined constants. 
These need to be fixed by other means. We do this explicitly, employing two alternative methods.
This results in a novel compact formula for the box integral valid in all kinematic regions of Minkowski space.
\end{minipage}
\end{center}

\newpage
\tableofcontents

\bigskip
\noindent\hrulefill
\smallskip

\section{Introduction}

The AdS/CFT correspondence continues to be one of the most inspiring and most cited conjectures of contemporary mathematical physics. Its structural deepness is e.g.\ seen in the quantum integrability of its free-AdS-strings/planar-Yang-Mills-theory limit.%
\footnote{The best overall review collection is still \cite{Beisert:2010jr}, even though there have been many further developments.}
 While being equally conjectural from a rigorous, non-perturbative point of view, integrability has already brought a large number of quantities on both the string as well as the gauge theory side under its spell, often leading to exact results that could not have been computed otherwise: String spectrum/scaling dimensions, Wilson loops and defect lines, various correlation functions, scattering amplitudes, and much more.

Integrability of free/planar AdS/CFT is arguably easier to see on the string side, where the supersymmetric string sigma model\footnote{Note, however, that the exact quantisation of superstrings on AdS$_5 \times$ S$^5$ still has not been achieved.}
is essentially an integrable two-dimensional quantum field theory. A systematic understanding (and even definition) of quantum integrability on the side of the four-dimensional ${\cal N}=4$ Super-Yang-Mills theory (SYM) seems to be a bigger challenge.
What is well-known, however, is that this model's integrable structure appears for absolutely any quantity one chooses to study in perturbation theory. The most successful and complete understanding has been reached for the computation of two-point functions, where a certain type of long-range spin chain emerges. In many cases, it even allows for deriving all-loop results. Interestingly, integrability has even been
`detected' on the level of individual tree- and loop-level Feynman diagrams. Here it usually appears in the form of {\it Yangian invariance}, see \cite{Loebbert:2016cdm,Ferro:2018ygf} for an introduction. The simplest case beyond tree level is the one-loop box diagram, which appears in the computations of four-point correlation functions as well as four-particle scattering amplitudes. In its Euclidean version, it reads
\begin{equation}
I_4^{\rm{E}}(x_1,x_2,x_3,x_4)=\frac{1}{\pi^2}\int
_{\mathbb{R}^4} 
\frac{\dd^4 x_0}{x_{01}^2 x_{02}^2 x_{03}^2 x_{04}^2},
\label{eq:BoxEuclidean}
\end{equation}
where $x_{jk}^2=(x_j-x_k)^2=\sum_{\mu=1}^4 (x_j^\mu-x_k^\mu)^2$ and $x_1,\dots, x_4$ are pairwise distinct points in $\mathbb{R}^4$. In this case the integral is finite, and has been computed a long time ago.
The most succinct way to explicitly write down the result employs its {\it (dual) conformal symmetry}: the function 
$
\hat\phi^E=x_{13}^2x_{24}^2I_4^E
$
 is invariant under arbitrary translations, four-dimensional rotations, scale transformations, as well as special conformal transformations.
Defining the a priori indistinguishable pair of conformal variables $z, \bar z$ through the two independent cross-ratios of the four points as
\begin{equation}
z\bar z =u=\frac{x_{12}^2 x_{34}^2}{x_{13}^2 x_{24}^2},
\qquad
(1-z)(1-\bar z)
=v
=
\frac{x_{14}^2 x_{23}^2}{x_{13}^2 x_{24}^2},
\label{eq:Definitionzzbar}
\end{equation}
the new function 
$
\phi^{\rm{E}}=  (z-\bar z)\,\hat \phi^E
$
depends only on $z, \bar z$ and may be written as
\begin{equation}
\phi^{\rm{E}}(z,\bar z)=
2\text{Li}_2(z) - 2\text{Li}_2(\bar z) + \brk*{\log z+\log \bar z} \brk1{\log ({1-z})-\log({1-\bar z})}.
\label{eq:hatphiEuclidean}
\end{equation}
In Euclidean space $z$ and $\bar z$ are necessarily related by complex conjugation.
Hence, $\phi^{\rm{E}}$ in \eqref{eq:hatphiEuclidean} is actually a function of the single complex variable $z$. It is a version of the dilogarithm function due to Bloch and Wigner, see \cite{Zagier:2007knq} for an in-depth discussion of it analytic properties.
As was demonstrated in detail in \cite{Chicherin:2017frs, Loebbert:2019vcj}, it is perfectly possible to derive this result from integrability. First, one shows that, infinitesimally, the dual conformal invariance of \eqref{eq:BoxEuclidean} is part of an even bigger symmetry: Yangian invariance. Secondly, one puts the latter to good use and derives a system of partial differential equations for 
$\hat \phi^{\rm{E}}(z,\bar z) $. These are easily solved, and lead to the following complete, fundamental system of linearly independent solutions for~${\phi}^\textrm{E}$: 
\begin{align}
f_1 &= 1, 
\label{eq:f1}
\\
f_2 &= \log{z}-\log{\bar z},
\label{eq:f2}
\\
f_3 &= \log ({1-z})-\log({1-\bar z}).
\label{eq:f3}
\\
f_4 &=2\text{Li}_2(z) - 2\text{Li}_2(\bar z) + \brk*{\log z+\log \bar z} \brk1{\log ({1-z})-\log({1-\bar z})}.
\label{eq:f4}
\end{align}
Thirdly, the correct linear combination is determined by using the permutation symmetry of the four points in Euclidean space, and the 
{\it star-triangle relation}.%
\footnote{Using the star-triangle relation required a generalised form of the box with generic propagator powers \cite{Loebbert:2019vcj}.}
 The latter is essentially a version of the Yang-Baxter equation, and therefore also based on integrability. One then indeed finds $\phi^{\rm{E}} = f_4$, i.e.\ exactly \eqref{eq:hatphiEuclidean}.

Now it is important to stress that AdS/CFT ultimately requires the consideration of quantum field theory in {\it Minkowski space}, as opposed to its `Wick-rotated' Euclidean version. The reason is subtly but stringently built on the theory of Lie superalgebras: The symmetry algebra of AdS/CFT is $\mathfrak{psu}(2,2|4)$. This includes as a subalgebra the conformal algebra of Minkowski space $\mathfrak{su}(2,2)\simeq \mathfrak{so}(2,4)\supset \mathfrak{so}(1,3)$, while for Euclidean space we have $\mathfrak{so}(4)\subset \mathfrak{so}(1,5) \simeq\mathfrak{su}^*(4)$. The latter algebra $\mathfrak{su}^*(4)$ is known to {\it not} be a subalgebra of $\mathfrak{psu}(2,2|4)$, nor of any other real form of the complexified Lie algebra $\mathfrak{psl}(4|4)$. In Minkowski space the integral \eqref{eq:BoxEuclidean} has to be replaced by 
\begin{equation}
I_4(x_1,x_2,x_3,x_4)=\frac{1}{i\pi^2}\int_{\mathbb{R}^{1,3}}  \frac{\dd^4 x_0}{(x_{01}^2 +i\epsilon)(x_{02}^2 +i\epsilon)(x_{03}^2 +i\epsilon)(x_{04}^2 +i\epsilon)},
\label{eq:BoxDefinition}
\end{equation}
where $x_{jk}^2=(x_j-x_k)^2=(x_j^0-x_k^0)^2-\sum_{l=1}^3 (x_j^l-x_k^l)^2$, $x_1,\dots, x_4$ are now four points in $\mathbb{R}^{1,3}$, and the Feynman-$i \epsilon$ prescription has to be implemented.%
\footnote{A technical remark: Here, with $\epsilon >0$, the variables $x_k$ of the box integral are region momenta. Of course, the box integral also appears for space-time Feynman diagrams, where we should replace $i \epsilon \rightarrow -i \epsilon$. In this case, `dual' conformal symmetry is replaced by ordinary conformal symmetry in space-time.} Now pairwise distinction of the four points no longer suffices for finiteness. Instead, it is necessary and sufficient to demand $x_{jk}^2 \neq 0$ for all $j,k$ with $j \neq k$. This integral has been calculated several times and studied from many different perspectives \cite{tHooft:1978jhc,Denner:1991qq,Usyukina:1992jd,Duplan_i__2002,hodges2010box}.
Dual conformal invariance holds only locally in this case, in particular certain large special conformal transformations can change the value of the integral~\cite{Corcoran:2020akn}. 
As before, one defines a pair of conformal variables $z, \bar z$ by \eqref{eq:Definitionzzbar}. However, depending on the signs of the individual $x_{jk}^2$, the variables $z, \bar z$ may now also become independently real, as well as complex conjugate pairs. In both cases the Euclidean result \eqref{eq:hatphiEuclidean} does not necessarily hold anymore. 
While the correct procedure to obtain the full result via analytic continuation in the complexified kinematical invariants (i.e.\ in our case, the set of six $x_{jk}^2$)
 is well-known in principle, it is quite 
involved in practice. 

A complete analysis of the box integral \eqref{eq:BoxDefinition} in all kinematic regions was recently performed in \cite{Corcoran:2020akn}. Interestingly, the result suggests that in all regions the Minkowski integral is still given by a specific linear combination of the four-dimensional solution space \eqref{eq:f1}--\eqref{eq:f4} of the Yangian bootstrap, where now all four functions are needed. This is encouraging, and naturally leads to an important question: {\it Is the Minkowski box integral} \eqref{eq:BoxDefinition} {\it also completely determined by integrability, just like its Euclidean precursor} \eqref{eq:BoxEuclidean}{\it ?}

In this paper, we will {\it almost} answer this question: By using Yangian invariance plus a few further discrete symmetries of the Minkowski box integral, we are able to severely constrain the final answer. Note that there are 64 kinematic regions, labeled by the signs of the six kinematical invariants $x_{jk}^2$, and therefore a priori $64 \times 4=256$ undetermined parameters, as each sector comes with a specific linear combination of the four-dimensional solution space \eqref{eq:f1}--\eqref{eq:f4}. We end up with twelve parameters that are currently undetermined by our use of integrability and symmetry.
These still need to be fixed by other means. Here we provide two alternative methods: analytic continuation in the $x_{jk}^2$, and evaluation for fixed spacetime configurations. Finally we present a compact formula, cf.\ \eqref{eq:Boxans}, from which the result for the box integral can be easily read off in all kinematic regions.

\section{The Box and its Symmetries}

The precise functional form of $I_4$ depends on the \emph{kinematic region} $R$, which is characterised by the sign vector of the six Mandelstam invariants, cf.\ \cite{Corcoran:2020akn}:
\begin{equation}
R= \sign \signblocktextbiggerbr{x_{12}^2, x_{34}^2}{x_{23}^2,x_{14}^2}{ x_{13}^2, x_{24}^2}.
\label{eq:signblock}
\end{equation}
Note that this notation underlines the different roles of the three rows in the above sign block, cf.\ the definition \eqref{eq:Definitionzzbar} of the conformal cross-ratios.
In the Euclidean region we have $x_{ij}^2<0$, and so $R=\signblockbr{--}{--}{--}$. In this region the box is invariant under all permutations of its external legs, which implies that it can be fully bootstrapped using its Yangian and permutation symmetries to be proportional to \emph{one} of the above Yangian invariants, see \cite{Loebbert:2019vcj}:
\begin{equation}
{\phi}\brk*{\signblock{--}{--}{--}}= f_4(z,\bar z).
\end{equation} 
Here we have
\begin{equation}
{\phi}=(z-\bar{z})\hat\phi=(z-\bar{z})x_{13}^2x_{24}^2I_4.
\end{equation}
This poses the question for the meaning of the remaining Yangian invariants $f_1$, $f_2$ and $f_3$.
Clearly, at least parts of the permutation symmetries are broken once the sign vector $R$ takes a less symmetric form than in the Euclidean region, and thus, imposing permutation symmetries will be less restrictive. It is the aim of this paper to extend the Yangian bootstrap from the Euclidean region discussed in \cite{Loebbert:2019vcj}, to all $2^6=64$ kinematic regions of the box in Minkowski space. In order to set the stage for this analysis, let us briefly discuss the different symmetries of the box integral \eqref{eq:BoxDefinition}.
For this purpose it is useful to also give the Feynman parametrisation of this integral, which takes the form 
\begin{equation}
 I_4=\int_{0}^{1}\dd^4y\frac{\delta(1-y_1-y_2-y_3-y_4)}{(\sum_{i<j}y_iy_jx_{ij}^2+i\epsilon)^2}.
\label{eq:FeynmanParam}
\end{equation}
We note that this representation is already well-defined beyond physical kinematics,%
\footnote{Physical kinematics are those which can be realised from four-point configurations in Minkowski space.} and so can be viewed as an analytic continuation of \eqref{eq:BoxDefinition}. In the following sections we will constrain the box integral for arbitrary real kinematics, which can then be restricted to phyiscal kinematics.

\paragraph{Yangian Symmetry.}
The box integral is the simplest member of a family of fishnet Feynman integrals that are invariant under the infinite-dimensional Yangian algebra \cite{Chicherin:2017frs,Chicherin:2017cns}. This algebra is spanned by two finite sets of level-zero and level-one generators with local and bi-local representations on the external legs of the Feynman graphs, respectively:
\begin{equation}
\gen{J}^a = \sum_{j=1}^n \gen{J}_{j}^a, 
\qquad\qquad
\levo{J}^a=\half f^a{}_{bc}\sum_{j<k=1}^n \gen{J}_{j}^c \gen{J}_{k}^b+ \sum_{j=1}^n \Eval_j \gen{J}_{j}^a.
\end{equation}
The level-zero generators are given by the ordinary conformal Lie algebra generators that act as differential operators on the variables $x_j$:
\begin{align}
\gen{P}^{\mu}_j &= -i \partial_{x_{j}}^{\mu}, &
\gen{L}_j^{\mu \nu} &= i x_j^{ \mu} \partial_{x_{j}}^{ \nu} 
	- ix^{ \nu}_j \partial_{x_{j}}^{\mu}, \nonumber \\
\gen{D}_j &= -i x_{j \mu} \partial_{x_j}^{ \mu} - i \Delta_j , &
\gen{K}^{\mu}_j &= -i \brk*{ 2 x_j^{\mu} x_j^{ \nu} 
		- \eta ^{\mu  \nu} x_j ^2 } \partial _{ \nu} 
	- 2i \Delta_j x_j^{\mu} .
\label{eq:level0gens}
\end{align}
Also the level-one generators are constructed from these generator densities. The level-one momentum generator for instance takes the form
\begin{equation}
\gen{\widehat P}^{\mu} = \sfrac{i}{2} \sum_{j<k=1}^n 
	\left(\gen{P}_j^\mu \gen{D}_k + \gen{P}_{j\nu} \gen{L}_k^{\mu\nu} 
		- (j\leftrightarrow k)\right) 
	+ \sum_{j=1}^n \Eval_j \gen{P}_j^{\mu}. 
\end{equation}
The box integral is annihilated by all level-zero and level-one generators for scaling dimensions $\Delta_j=1$ and evaluation parameters
\begin{equation}
\Eval_j=\brk*{+\sfrac{3}{2},+\half,-\half,-\sfrac{3}{2}}_j,
\qquad j=1,\dots,4.
\end{equation}
The resulting partial differential equations are highly constraining as discussed in \Secref{sec:constraints}.


\paragraph{Permutation Symmetry.}

As is clear from \eqref{eq:BoxDefinition}, the box integral $I_4$ is invariant under all permutations $\sigma\in S_4$ of the external points
\begin{equation}
\sigma:(x_1,x_2,x_3,x_4)\rightarrow(x_{\sigma(1)},x_{\sigma(2)},x_{\sigma(3)},x_{\sigma(4)}).
\end{equation}
This results in the function $\phi(z,\bar{z})=(z-\bar{z})x_{13}^2x_{24}^2I_4$ transforming very simply under permutations. $\sigma$ generates transformations of the conformal invariants $z\rightarrow h_{\sigma}(z), \bar{z}\rightarrow h_{\sigma}(\bar{z})$, as well as a transformation of the kinematic region $R\rightarrow R_\sigma$. This leads to the equation
\begin{equation}
\phi(z,\bar{z}|R)=\sign(\sigma)\sp\phi(h_\sigma(z),h_\sigma(\bar{z})|R_\sigma).
\label{eq:permutation}
\end{equation} 
The full behaviour of $z$, $\bar{z}$ and $R$ under permutations is outlined in \Tabref{table:permutations} of \Appref{appendix:details}.


\paragraph{Shuffling Symmetry.}
There is a large redundancy in the space of kinematic regions $R$, which we will denote as `shuffling' symmetry.\footnote{This of course has nothing to do with the shuffle product.} The box integral has a symmetry under simultaneous exchange of two pairs of kinematic variables\footnote{Note that using the Mellin representation one can show invariance under individual exchanges of these variables, however, this is not needed for our argument.}
\begin{equation}
\begin{array}{c} x_{12}^2\leftrightarrow x_{34}^2 \\ x_{23}^2\leftrightarrow x_{14}^2 \end{array},\spac \begin{array}{c} x_{12}^2\leftrightarrow x_{34}^2 \\ x_{13}^2\leftrightarrow x_{24}^2 \end{array},\spac \begin{array}{c} x_{23}^2\leftrightarrow x_{14}^2 \\ x_{13}^2\leftrightarrow x_{24}^2 \end{array},
\end{equation}
which is clear from the representation \eqref{eq:FeynmanParam}. This means that given a kinematic region $R$, represented by a sign block \eqref{eq:signblock}, the box integral is invariant under the operation of swapping the signs in exactly two of the rows of the sign block, for example $\signblockbr{-+}{++}{+-}\rightarrow \signblockbr{+-}{++}{-+}$. If $R$ and $R'$ are related by such a shuffle, then we have
\begin{equation}
\phi(z,\bar{z}| R)=\phi(z,\bar{z}| R').
\end{equation}

\paragraph{Conjugation Symmetry.} Simultaneous reversal of the signs of the kinematic invariants $x_{ij}^2$ is equivalent to complex conjugation of the integral. This is most easily expressed in terms of the function $\hat{\phi}=\phi/(z-\bar{z})$:
\begin{equation}
\hat{\phi}(z,\bar{z}| R)=\hat{\phi}(z,\bar{z}| -R )^*,
\label{eq:conjugation}
\end{equation}
which is easily proven from the representation \eqref{eq:FeynmanParam}.
\section{Constraints from Symmetries}
\label{sec:constraints}
In this section we impose the constraints of the above symmetries. We will find that this reduces the computation of the box integral in all kinematic regions to fixing only twelve constant parameters. These will be determined in \Secref{sec:FixConstants}.

\paragraph{Yangian Symmetry.}

Dual conformal (level-zero) invariance implies that the box integral can be written in the form
\begin{equation}
I_4=\frac{1}{x_{13}^2x_{24}^2}\hat\phi(z,\bar z),
\end{equation}
with $z,\bar z$ defined in \eqref{eq:Definitionzzbar}. Yangian level-one invariance $\levo{J}^aI_4=0$ in addition implies the  two differential equations
\cite{Loebbert:2019vcj}
\begin{equation}
0=\brk[s]*{D_j(z)-D_j(\bar z)}\hat\phi(z,\bar z),
\qquad j=1,2,
\label{eq:level1PDEs}
\end{equation}
where we employ the differential operators
\begin{align}
D_1(z)&=z(z-1)^2\partial_z^2+(3z-1)(z-1)\partial_z+z,
\\
D_2(z)&=z^2(z-1)\partial_z^2+(3z-2)z\partial_z+z.
\end{align}
In  \cite{Loebbert:2019vcj} the Yangian differential equations \eqref{eq:level1PDEs} have been solved to find that the general solution is given by a linear combination of the four basis functions \eqref{eq:f1}--\eqref{eq:f4} with (piecewise) constant coefficients.
In the present paper we choose to work with a slightly different basis of these four functions given by
\begin{align}
g_1&=2\text{Li}_2(z) - 2\text{Li}_2(\bar z) + \brk*{\log z+\log \bar z} \brk1{\log ({1-z})-\log({1-\bar z})},
\label{eq:gbasis1}
\\
g_2&=\log(z)-\log(\bar{z}),
\label{eq:gbasis2}
\\
g_3&=\log(1-z)-\log(1-\bar{z}),
\label{eq:gbasis3}
\\
g_4&=\log(1-\sfrac{1}{z})-\log(1-\sfrac{1}{\bar{z}}).
\label{eq:gbasis4}
\end{align}
This basis is motivated by the fact that the three functions $g_2,g_3,g_4$ can be understood as single discontinuities of the highest transcendentality function $g_1$.%
\footnote{Note on the other hand that the function basis $f_1,f_2,f_3,f_4$ as given in \eqref{eq:f1}--\eqref{eq:f4} reveals that a transcendentality-zero solution to the Yangian constraints exists. The function $g_4$ can be understood as a single discontinuity of $g_1$ about $z=\infty$, with $\bar{z}$ fixed.} 
Note that it was demonstrated in \cite{Chicherin:2017cns} that the Yangian invariance equations do not change if we replace a $1/x_{0j}^2$ propagator by $\delta(x_{0j}^2)$. This shows that the solution space of these equations does also contain the cuts of the box integral, which in turn relate to its discontinuities, cf.\ \cite{Abreu:2014cla} and \Secref{sec:FixConstants}.

There are a few technicalities to mention. In Euclidean space the conformal invariants are always constrained via $\bar{z}=z^*$. However, in Minkowski space $z$ and $\bar{z}$ can also be independent real numbers. The possible range of $z$ and $\bar{z}$ depends on the kinematic region, and is summarised for arbitrary real kinematics in \Tabref{table:zzbrange} of \Appref{appendix:details}. The functions $g_j$ have branch cuts on various intervals of the real axis, which are fixed after specifying the usual branch of the logarithm on the negative real axis. In order to consistently define the functions $g_j$ for all possible values of $z,\bar{z}$, we regularise them according to
\begin{align}
g^{-}_j(z,\bar{z}) &=g_j(z-i\delta,\bar z+i\delta),\spac j=1,2,3,4,
\label{eq:gbasisreg}
\end{align}
where $\delta$ is an infinitesimal positive real number. We similarly define regularised functions $g_j^{+}$ identically to the above, but with the replacement $ \delta \rightarrow - \delta$. Note that such regularisations break the antisymmetry of the functions $g_j$ in $z$ and $\bar{z}$. As such, we explicitly specify $z$ and $\bar{z}$ in terms of the cross-ratios $u$ and $v$ from \eqref{eq:Definitionzzbar} as
\begin{align}
    z&=\sfrac{1}{2}(1+u-v)+ \sfrac{1}{2}\sqrt{(1-u-v)^2-4 u v},
    \nonumber\\
    \bar{z}&=\sfrac{1}{2}(1+u-v)- \sfrac{1}{2}\sqrt{(1-u-v)^2-4 u v},
    \label{eq:zzbarsol}
\end{align}
so that in particular $z\geq \bar{z}$ when $z,\bar{z}\in\mathbb{R}\setminus\{0,1\}$ and $\text{Im}(z)>0$ when $z\in\mathbb{C}\setminus \mathbb{R}, \bar{z}=z^*$. Then the exchange $z\leftrightarrow\bar{z}$ is equivalent to $g_j^{+}\leftrightarrow -g_j^{-}$. Note that fixing $z$ and $\bar{z}$ as in \eqref{eq:zzbarsol} can lead to them `swapping' after a permutation $\sigma$. For example, $\sigma=(13)$ generates the transformation of conformal invariants $z\rightarrow z'=1-\bar{z}$ and $\bar{z}\rightarrow \bar{z}'=1-z$. In general, whether $z$ and $\bar{z}$ swap after a permutation depends on both the range of $z$ and $\bar{z}$ and the permutation. Since we are imposing permutation symmetry on our final function $\phi$, we allow for the appearance of both functions $g_j^{+}$ and $g_j^{-}$ in our ansatz derived from Yangian invariance. In the end we can always express the functions $g_{j}^{\mp}$ in terms of $g_{j}^{\pm}$ to get an expression for the integral in terms of just four regularised functions $g_j^{-}$ or $g_j^{+}$, see \Appref{appendix:details}. Our final ansatz is
\begin{equation}
 \phi(R)=\sum_{i=1}^4\left(\alpha_i^R g_{i}^{-}+\beta_i^R g_{i}^{+}\right),
 \label{eq:Yangsatz}
\end{equation}
where $\alpha_i^R$ and $\beta_i^R$ are complex numbers depending on the kinematic region $R$. In total there are $64\times 4$ = 256 constants $\alpha_i^R$ to fix. $\beta_i^{R}$ can be expressed in terms of $\alpha_i^R$ using the transition matrix \eqref{eq:transition}.


\paragraph{Permutations, Shuffles, and Conjugation.}
A priori we have $2^6=64$ functions ${\phi}(R)$ to fix, one for each of the possible kinematic regions $R$. However, permutation, shuffling, and conjugation symmetry already give very large constraints on the linear combination \eqref{eq:Yangsatz}. Under these three operations there are \emph{six} equivalence classes of sign blocks. We list a representative from each equivalence class:
\begin{align}
R_1&=\signblockbiggerbr{--}{--}{--},&  R_2&=\signblockbiggerbr{-+}{-+}{-+}, & R_3&=\signblockbiggerbr{-+}{++}{++},
&
 R_4&= \signblockbiggerbr{--}{-+}{-+},& R_5&=\signblockbiggerbr{--}{++}{++},& R_6&=\signblockbiggerbr{-+}{++}{--}.
\end{align}
Using shuffling symmetry, we can identify $-+\sim +-$ in any row of the sign block whenever there is at least one other row invariant under swapping ($--$ or $++$). We will always choose the order $-+$ when possible; the only time it is not possible is for the signature $-R_2=\signblockbr{+-}{+-}{+-}$. These considerations already restrict the number of independent signatures to $3^3+1=28$. These remaining 28 signatures organise themselves into six equivalence classes $\Lambda_1,\Lambda_2,\dots,\Lambda_6$ under permutations and conjugations. $\Lambda_1$ contains $R_1$ and $-R_1$, and $\Lambda_2$ contains $R_2$ and $-R_2$. The remaining equivalence classes each contain six signatures, for example 
\begin{equation}
\Lambda_3=\left\{\signblockbiggerbr{-+}{++}{++},\signblockbiggerbr{++}{-+}{++},\signblockbiggerbr{++}{++}{-+},\signblockbiggerbr{-+}{--}{--},\signblockbiggerbr{--}{-+}{--},\signblockbiggerbr{--}{--}{-+}\right\}.
\end{equation}
The full set of $\Lambda_i$ is given explicitly in \eqref{eq:equivalence}. When the box integral is known in the representative kinematic region $R_i$, it can be deduced for each of the remaining signatures in $\Lambda_i$ using \eqref{eq:permutation} and \eqref{eq:conjugation}. Therefore we only need to fix $4\times 6=24$ constants, 4 for each of the $R_i$. We can eliminate some of these constants by using the fact that some of the $R_i$ are invariant under a subgroup of $S_4$.
\paragraph{Region $ R_1$.} 
$R_1$
is fully invariant under permutations. For example, invariance under the transposition $(12)$ gives a constraint on the ansatz \eqref{eq:Yangsatz}:
\begin{equation}
\alpha_i^{R_1} g_{i}^{-}(z,\bar{z})+\beta_i^{R_1} g_{i}^{+}(z,\bar{z})=-\alpha_i^{R_1} g_{i}^{+}(1-z,1-\bar{z})-\beta_i^{R_1} g_{i}^{-}(1-z,1-\bar{z}),
\label{eq:A1constrain}
\end{equation}
where summation over $i$ is assumed. In \eqref{eq:g1perm}--\eqref{eq:g4perm} we list the behaviour of the regularised functions $g_i^\pm$ under permutations. Only $g_1$ and $g_4$ are compatible with the functional equation \eqref{eq:A1constrain}, which forces $\alpha_2^{R_1}=\beta_2^{R_1}=\alpha_3^{R_1}=\beta_3^{R_1}=0$. Furthermore, invariance of $R_1$ under the transposition $(14)$ forces $\alpha_4^{R_1}=\beta_4^{R_1}=0$. Therefore \eqref{eq:Yangsatz} reduces to
\begin{equation}
 \phi(R_1)=\alpha_1^{R_1} g_{1}^{-}+\beta_1^{R_1}g_{1}^{+}.
 \label{eq:A1phi1}
\end{equation}
For the possible values of $z,\bar{z}$ in the kinematic region $R_1$ (see \Tabref{table:zzbrange}), we could use \eqref{eq:transition} for example to deduce $g_1^{+}=g_1^{-}$. Therefore we can rewrite \eqref{eq:A1phi1} as
\begin{equation}
 \phi(R_1)=a_1g_1^{-}=a_1g_1^{+},
\end{equation}
where $a_1= \alpha_1^{R_1}+\beta_1^{R_1}\in\mathbb{C}$. The box integral for the other signature in $\Lambda_1$ can be calculated using \eqref{eq:conjugation}:
\begin{equation}
 \phi(-R_1)=a_1^*g_1^{+}=a_1^*g_1^{-}.
 \label{eq:mR1}
\end{equation}
Therefore, for the equivalence class $\Lambda_1$, there is only a single constant $a_1$ left to be fixed. We leave most of the details for the remaining representative signatures $R_i$ to \Appref{appendix:details}, and merely state the results.
\paragraph{Region $ R_2$.} $R_2$
is also fully invariant under permutations, and similarly to $R_1$ we constrain
\begin{equation}
\phi(R_2)=a_2g_1^{-}=a_2g_1^{+}.
 \end{equation}
 ${\phi}(-R_2)$ is calculated analogously to \eqref{eq:mR1}. Therefore, in the equivalence class $\Lambda_2$ there is also just a single constant $a_2$ to fix.
\paragraph{Region $ R_3$.} $R_3$
is not completely invariant under permutations, and so we expect the functional form of ${\phi}$ to be less restricted. We do have invariance under the transposition $(12)$ however, which restricts the linear combination \eqref{eq:Yangsatz} to
\begin{equation}
{\phi}(R_3)=a_3 g_{1}^{-}+2\pi i c_3g^{-}_3=a_3g_{1}^{+}+2\pi i(c_3+a_3)g^{+}_3,
\label{eq:A3phi}
\end{equation}
so that in $\Lambda_3$ there are two constants $a_3$ and $c_3$ to fix. We write the coefficient of $g_3^-$ as $2\pi i c_3$ for later convenience. We give the derivation of \eqref{eq:A3phi} in  \Appref{appendix:details}. The box integral for the remaining signatures in $\Lambda_3$ can then be derived using \eqref{eq:permutation} and \eqref{eq:conjugation}.
\paragraph{Region $ R_4$.} $R_4$
 is also invariant under the transposition (12). Analogously to $R_3$ this leads to 
\begin{equation}
    \phi(R_4)=a_4 g_{1}^{-}+2\pi ic_4g^{-}_3=a_4g_{1}^{+}+2\pi i(c_4+a_4)g^{+}_3,
\end{equation}
so that there are two constants $a_4$ and $c_4$ to fix in $\Lambda_4$.
\paragraph{Region $ R_5$.} $R_5$ is also
invariant under the transposition (12), which leads to 
\begin{align}
\phi(R_5)
&=a_5g_1^{-}+2\pi i c_5g_3^{-}
\nonumber\\
&=a_5g_1^{+}+2\pi i c_5g_3^{+}+4\pi i c_5\theta_1\bar{\theta}_1(g_2^+-g_3^++g_4^+),
\label{eq:phiR5}
\end{align}
where $\theta_1= \theta(z-1)$ and $\bar{\theta}_1= \theta(\bar{z}-1)$. Therefore there are two constants $a_5$ and $c_5$ to fix in $\Lambda_5$.
Note the appearance of the theta functions in the second line of \eqref{eq:phiR5} renders the $g_i^-$ basis slightly more natural.
\paragraph{Region $ R_6$.} 
The region $R_6$
has no symmetry under nontrivial permutations, and therefore we cannot derive a constraint as easily as in the previous cases. We thus have
\begin{align}
\phi(R_6)
&=a_6g^{-}_1+2\pi ib_6g^{-}_2+2\pi ic_6g^{-}_3+2\pi id_6g^{-}_4
\nonumber\\
&=\alpha_6g^{+}_1-2\pi id_6g^{+}_2+2\pi i\bar{c}_6 g^{+}_3-2\pi ib_6g^{+}_4,
\label{eq:A6phi1}
\end{align}
where $\bar{c}_6=a_6+b_6+c_6+d_6$. Therefore there are still four constants $a_6,b_6,c_6,d_6$ to fix in $\Lambda_6$.

\paragraph{Combined Symmetries.}
To summarise, combining the above symmetries yields the following form for the box integral that depends on twelve unfixed parameters:
\begin{alignat}{2}
\phi(z,\bar z|R)=
&+a_R g_1^{-}&&
\nonumber\\
&+2\pi i g_2^{-}\big(&&-c_3\theta\signblock{++}{-+}{++}+\hat{c}_3^*\theta\signblock{--}{-+}{--}-c_4\theta\signblock{-+}{--}{-+}+\hat{c}_4^*\theta\signblock{-+}{++}{-+}-c_5\theta\signblock{++}{--}{++}+c_5\theta\signblock{--}{++}{--}
\nonumber\\
&&&+b_6\theta\signblock{-+}{++}{--}+d_6\theta\signblock{-+}{--}{++}-c_6\theta\signblock{++}{-+}{--}-d_6\theta\signblock{--}{++}{-+}-b_6\theta\signblock{++}{--}{-+}-c_6\theta\signblock{--}{-+}{++}\big)
\nonumber\\
&+2\pi i g_3^{-}\big(&&+c_3\theta\signblock{-+}{++}{++}-\hat{c}_3^*\theta\signblock{-+}{--}{--}+c_4\theta\signblock{--}{-+}{-+}-\hat{c}_4^*\theta\signblock{++}{-+}{-+}+c_5\theta\signblock{--}{++}{++}-c_5\theta\signblock{++}{--}{--}
\nonumber\\
&&&+c_6\theta\signblock{-+}{++}{--}+ c_6\theta\signblock{-+}{--}{++}-b_6\theta\signblock{++}{-+}{--}+b_6\theta\signblock{--}{++}{-+}+d_6\theta\signblock{++}{--}{-+}-d_6\theta\signblock{--}{-+}{++}\big)
\nonumber\\
&+2\pi i g_4^{-}\big(&&-\hat{c}_3\theta\signblock{++}{++}{-+}+c_3^*\theta\signblock{-+}{--}{--}-\hat{c}_4\theta\signblock{-+}{-+}{--}+c_4^*\theta\signblock{-+}{-+}{++}-c_5\theta\signblock{++}{++}{--}+c_5\theta\signblock{--}{--}{++}
\nonumber\\
&&&+d_6\theta\signblock{-+}{++}{--}+b_6\theta\signblock{-+}{--}{++}+d_6\theta\signblock{++}{-+}{--}-\bar{c}_6\theta\signblock{--}{++}{-+}-\bar{c}_6\theta\signblock{++}{--}{-+}+b_6\theta\signblock{--}{-+}{++}\big).
\label{eq:mastermp}
\end{alignat}
Here $a_R\in\{a_i,a_i^*\}$ depends on the kinematic region, e.g.\ $a_{R_2}=a_2$, and we abbreviate $\hat{c}_i= c_i+a_i$. 
Moreover, we have introduced the above theta-functions such that 
\begin{equation}
\theta{R'}(R)=
\begin{cases}
1 \quad \text{if}\quad  R'=R,
\\ 0\quad \text{otherwise}.
 \end{cases}
 \label{eq:definitiontheta}
 \end{equation}
For instance we have
\begin{equation}
\theta\signblock{++}{-+}{++}\brk*{\signblock{++}{-+}{++}}=1,
\qquad\qquad
\theta\signblock{++}{-+}{++}\brk*{\signblock{-+}{--}{++}}=0.
\end{equation}
 Alternatively, we can express the above ansatz in terms of $g_i^{+}$ as given in \eqref{eq:altans}. 

\section{Fixing Constants}\label{section:fix}
\label{sec:FixConstants}
After exhausting the available symmetries of the box integral, we are left with twelve independent constants that remain to be fixed. While we are confident that in the future also these can be determined using integrability, for the moment we consider them as additional input that is obtained by some method of choice. For instance, we could calculate the box integral for a set of arbitrary numerical configurations in the relevant regions to fix these numbers. In \Appref{app:Fixingv2} we show how to obtain the remaining constants by evaluating the box integral for particular `double infinity' configurations. In the present section we explicitly demonstrate how to obtain the twelve parameters using analytic continuation. 

To connect the box integral in different kinematic regions, we note that it is always represented by the same Feynman parametrised integral \eqref{eq:FeynmanParam} which gives a natural analytic continuation beyond real kinematics. In particular, this tells us that away from its poles, \eqref{eq:FeynmanParam} is a continuous function of the $x_{ij}^2$. Hence, we can relate the value of the box integral in different regions by connecting them via paths in $x_{ij}^2$ space on which the integral is regular. Since the integral diverges at points where one of the $x_{ij}^2$ vanishes, to change the signature of the kinematics on a regular path, we will have to continue the function through the complex plane. In this process, for generic complex $x_{ij}^2$, $z$ and $\bar z$ will cross branch cuts of the function basis $g_i$. Carefully tracking the movement of $z$ and $\bar z$ and adding or subtracting the corresponding discontinuities will therefore allow us to deduce the functional representation of the box integral in any of the regions. 

As a practical definition of the discontinuity of a function we use
\begin{align}
\text{Disc}_{z=a} f(z) = \lim_{\epsilon\rightarrow 0}\left(f(\gamma(\epsilon)) - f(\gamma(1-\epsilon))\right),
\end{align} 
where $\gamma(t)$ is a complex contour that encircles the branch point $a$ once on a clockwise path and starts and ends at $z$, i.e.\ $\gamma(0)=\gamma(1)=z$.\footnote{In particular, this definition is equivalent to the prescription given in \cite{Bourjaily:2020wvq} where discontinuities are evaluated via line integrals around the branch points.} For branch cuts of $f$ on the real axis, this definition implies for $a,z\in \mathds{R}$ and $z$ on a branch cut starting at $a$:
\begin{align}
\text{Disc}_{z=a} f(z) = \lim_{\epsilon\rightarrow 0}\left(f(z\pm i \epsilon) - f(z \mp i \epsilon)\right),
\end{align}
Here the sign of the $i \epsilon$ depends on the ordering of $a$ and $z$. This expression is easy to evaluate and sufficiently general for the set of functions $g_i$. We choose the branch cuts of our function basis to lie on the real axis, which is consistent with taking the principal value of the appearing logarithm and dilogarithm functions. To be precise, we have listed the branch cuts in \Tabref{table:branchcuts}.
\begin{table}
\begin{center}
 \begin{tabular}{|c|c|} 
 \hline
 Function &  Branch cut in $z$, $\bar z$ \\ [0.5ex] 
 \hline

 $g_1$ & $(-\infty, 0]$, $[1,\infty)$\\

 \hline
$g_2$ & $(-\infty, 0]$\\

 \hline
$g_3$ & $[1,\infty)$\\

 \hline
$g_4$ & $[0,1]$\\

 \hline
\end{tabular}
\caption{Location of branch cuts for the function basis $g_i$.}
\label{table:branchcuts}
\end{center}
\end{table}\noindent 

For convenience, we explicitly note the non-vanishing discontinuities of the $g_i$ around their branch points 
\begin{align}
 \text{Disc}_{z=0} g_1 &=+ \text{Disc}_{\bar z=0} g_1 = +2\pi i g_3, &\text{Disc}_{z=1} g_1 &= +\text{Disc}_{\bar z=1} g_1 =- 2\pi i g_2, \notag\\
\text{Disc}_{z=0} g_2 &= -\text{Disc}_{\bar z = 0} g_2 = +2\pi i, & \text{Disc}_{z=1} g_3 &= -\text{Disc}_{\bar z = 1} g_3 = +2\pi i, \notag\\
\text{Disc}_{z=1} g_4 &= -\text{Disc}_{\bar z = 1} g_4 = -2\pi i, &
 \text{Disc}_{z=0} g_4 &= -\text{Disc}_{\bar z = 0} g_4 = -2\pi i.
\end{align}
In the Euclidean region $R_1 =\signblockbr{--}{--}{--}$ the remaining coefficient $a_1$ is fixed by the star-triangle relation for generic propagator powers \cite{Loebbert:2019vcj},  such that
\begin{align}
\phi(R_1) = g_1^{+}=g_1^{-},
\end{align}
i.e.\ $a_1 = 1$. We will use this region as a starting point of the paths leading into the five other equivalence classes. Since for real kinematics, $z$ and $\bar z$ are either real or a pair of complex conjugates in the Euclidean region (see \Tabref{table:zzbrange}), we can always set up an entirely real path in kinematics space that sends all $x_{ij}^2$ to $-1$ without picking up any discontinuities: any possible branch cut passage will happen simultaneously for $z$ and $\bar z$ and give cancelling contributions. 
Hence to connect regions where some of the $x_{ij}^2$ differ in sign, for simplicity we can restrict ourselves to paths of the form 
\begin{equation}
x^2_{ij}=e^{i \varphi_{ij}}.
\end{equation}
To ensure that we do not encounter any poles of the integrand of \eqref{eq:FeynmanParam} we further demand $0<\varphi_{ij}<\pi$, i.e.\ we always rotate the $x_{ij}^2$ through the upper half of the complex plane.
This prescription is inherited from the positive $i\epsilon$-shift in the original expression \eqref{eq:BoxDefinition} for the box integral, which in turn translates into a positive $i\epsilon$-shift of the $x_{ij}^2$ in the Feynman parametrisation \eqref{eq:FeynmanParam}, c.f.\ \cite{Abreu:2014cla}.
Then, the paths on which we analytically continue from a region $R_k$ to another region $R_l$, can be parametrised by

\begin{align}
\varphi_{ij} = \left\{\begin{array}{cl} 
0, &\text{ if } \text{sgn}_{k}(x_{ij}^2) =+ \text{sgn}_l(x_{ij}^2) = +1\\
\pi, &\text{ if } \text{sgn}_{k}(x_{ij}^2) =+ \text{sgn}_l(x_{ij}^2) = -1\\
\varphi, &\text{ if } \text{sgn}_{k}(x_{ij}^2) = -\text{sgn}_l(x_{ij}^2) =+ 1\\
\pi - \varphi, &\text{ if } \text{sgn}_{k}(x_{ij}^2) = -\text{sgn}_l(x_{ij}^2) = -1
\end{array}\right.
,
\end{align}
where $\varphi\in [0,\pi]$ and $\text{sgn}_k(x_{ij}^2)$ is the sign of $x_{ij}^2$ in the region $R_k$. 

\begin{figure}
\begin{center}
\begin{tabular}{c l c l c l c l }
a) & \includegraphicsbox[scale=0.63]{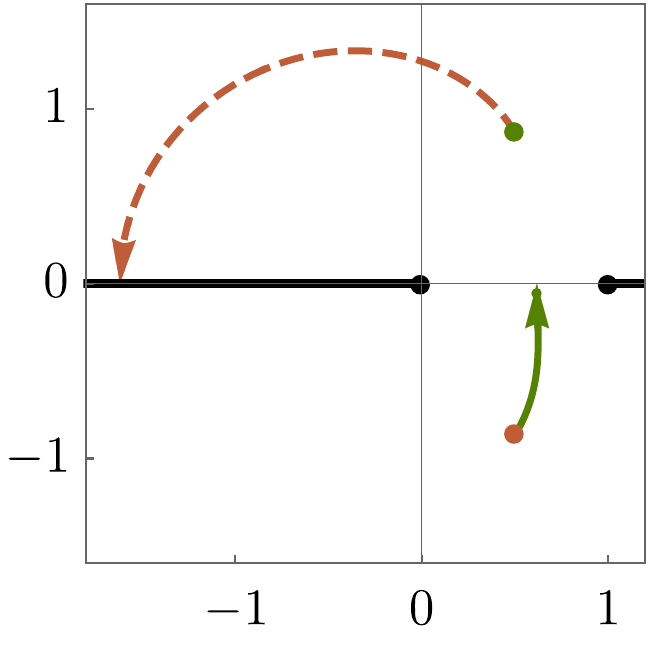}
&
b) & \includegraphicsbox[scale=0.63]{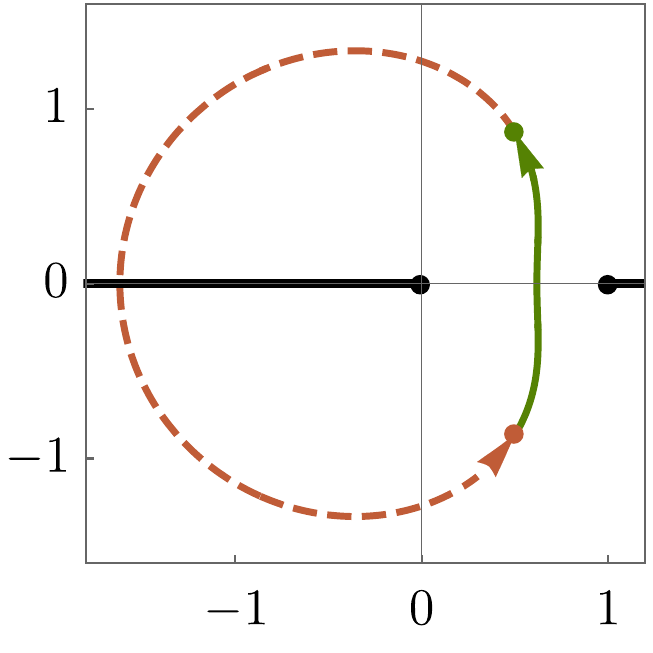}
&
c) & 
\includegraphicsbox[scale=0.63]{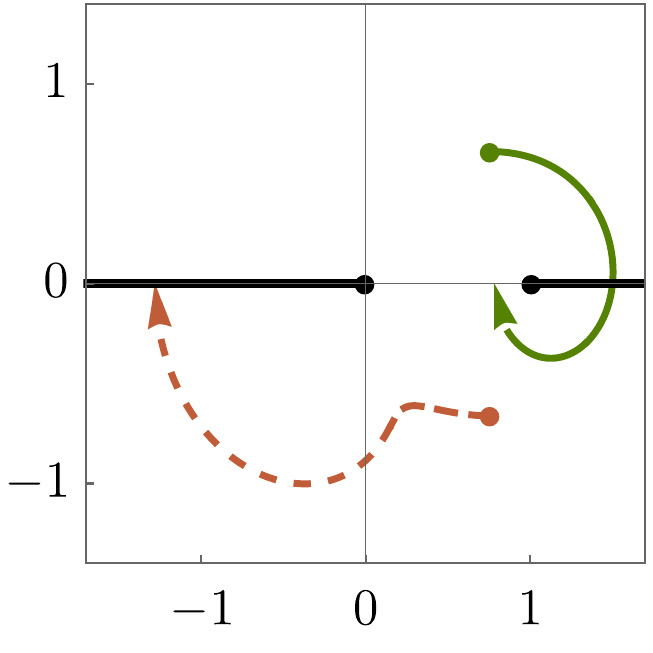}
\end{tabular}
\end{center}
\caption{The path of $z$ (solid green) and $\bar z$ (dashed red) under the continuation from $R_1$ to (a) $R_3$ or $R_4$, (b) $R_5$ and (c) $R_6$. In the last case we further introduce a factor of $1/2$, i.e.\ we set $x_{23}^2 = e^{i \varphi_{23}}/2$, to avoid the degenerate point $z=\bar z$. We denote the point where the paths begin by a dot in the respective color. Note that for (a) and (b) the branch cut of the square root in (\ref{eq:zzbarsol}) is passed immediately at the beginning of the path. Since away from the real line, $g_i^\pm(z,\bar z) = -g_i^\pm(\bar z, z)$ this does not change the result for $\phi$.  We also include the branch cuts of $g_1$ as solid black lines.
}
\label{Fig:RegionA2}
\end{figure}

\paragraph{Region $R_2$.} For $R_2=\signblockbr{-+}{-+}{-+}$ we have
\begin{align}
\varphi_{34} = \varphi_{14} = \varphi_{24} = \pi-\varphi
\end{align}
and all other $\varphi_{ij}$ equal to $\pi$. Hence, $u$ and $v$ are actually inert under this analytical continuation and so are $z$ and $\bar z$. Therefore, we find
\begin{align}
\phi(R_2) =  g_1^+=g_1^-,
\end{align}
fixing $a_2 = 1$.
\paragraph{Region $R_3$.} For $R_3=\signblockbr{-+}{++}{++}$ we have
\begin{align}
\varphi_{12} = \pi
\end{align}
as the only constant phase and all other equal to $\pi-\varphi$. The path that $z$ and $\bar z$ trace out under this continuation is shown in \Figref{Fig:RegionA2}a). Since the path for $\bar z$ ends on the negative real axis, the concrete expression in terms of $g_i$ depends on the regularisation procedure. For the $g_i^+$, the branch cut in $\bar z$ lies slightly above the negative real axis, whereas for $g_i^-$ it lies slightly below. Therefore, we introduce the operators $\theta^\pm$ that act according to
\begin{align}
\theta^\pm g_i^\pm &= g_i^\pm, & \theta^\pm g_i^\mp &= 0.
\end{align}
Then, we can compactly write the result in the region $R_3$ as
\begin{align}
\phi(R_3) = (1+\theta^+\text{Disc}_{\bar z=0})\phi(R_1) =  g_1^{-} = g_1^{+} + 2\pi i g_3^{+},
\end{align}
fixing $a_3 = 1+c_3 = 1$.
\paragraph{Region $R_4$.}  In order to move from $R_1=\signblockbr{--}{--}{--}$ to $R_4=\signblockbr{--}{-+}{-+}$, we choose
\begin{align}
\varphi_{14}=\varphi_{24}=\pi-\varphi,
\end{align}
and all other angles equal to $\pi$. Interestingly, this induces the very same path for $u$ and $v$ (and hence for $z$ and $\bar z$) as the continuation in the previous paragraph. Hence, we conclude
\begin{align}
\phi(R_4)=\phi(R_3) =  g_1^{-} = g_1^{+} + 2\pi i g_3^{+},
\end{align}
fixing $a_4 = 1+c_4 = 1$.
\paragraph{Region $R_5$.} For the region  $R_5=\signblockbr{--}{++}{++}$ we find
\begin{align}
\varphi_{13}=\varphi_{24}=\varphi_{23}=\varphi_{41}=\pi-\varphi,
\end{align}
and all other angles equal to $\pi$. From the $z$, $\bar z$ path shown in \Figref{Fig:RegionA2}b), we conclude
\begin{align}
\phi(R_5) = (1+\text{Disc}_{\bar z=0}) \phi(R_1) = g_1^\pm +2\pi i g_3^\pm,
\end{align}
independently of the regularisation.

\paragraph{Region $R_6$.} Finally, for $R_6=\signblockbr{-+}{++}{--}$ the non-constant phases are
\begin{align}
\varphi_{34}=\varphi_{23}=\varphi_{14}=\pi-\varphi.
\end{align}
The corresponding path for $z,\bar z$  in \Figref{Fig:RegionA2}c) implies
\begin{align}
\phi(R_6) &= (1-\theta^- \text{Disc}_{\bar z = 0})(1-\text{Disc}_{z=1}) \phi(R_1)\notag\\
&= g_1^{+}+2\pi i g_2^{+} = g_1^{-}+2\pi i(g_2^{-}-g_3^{-} +2\pi i)\notag\\
&= g_1^- - 2\pi i g_4^-,
\end{align}
fixing $a_6 = -d_6 = 1$ and $b_6=c_6=\bar{c}_6 = 0$.
\paragraph{Summary.}

In summary, we find 
\begin{equation}
a_R=1,\qquad
c_6=\bar{c}_6=b_6=0,\qquad
1+c_3=1+c_4=c_5=-d_6=1,\qquad
\end{equation}
such that the ansatz \eqref{eq:mastermp} yields the full result for the box integral:
\begin{align}
\phi(R)
=
&+ g_1^{-}
\nonumber\\
&+2\pi i g_2^{-}\brk*{+\theta\signblock{--}{-+}{--}+\theta\signblock{-+}{++}{-+}-\theta\signblock{++}{--}{++}+\theta\signblock{--}{++}{--}-\theta\signblock{-+}{--}{++}+\theta\signblock{--}{++}{-+}}
\nonumber\\
&+2\pi i g_3^{-}\brk*{-\theta\signblock{-+}{--}{--}-\theta\signblock{++}{-+}{-+}+\theta\signblock{--}{++}{++}-\theta\signblock{++}{--}{--}-\theta\signblock{++}{--}{-+}+\theta\signblock{--}{-+}{++}}
\nonumber\\
&+2\pi i g_4^{-}\brk*{-\theta\signblock{++}{++}{-+}-\theta\signblock{-+}{-+}{--}-\theta\signblock{++}{++}{--}+\theta\signblock{--}{--}{++}-\theta\signblock{-+}{++}{--}-\theta\signblock{++}{-+}{--}}.
\label{eq:Boxans}
\end{align}
For the benefit of the reader, we collect once more the definitions of the various terms in \eqref{eq:Boxans}. The regularised functions $g_j^-$ are given in \eqref{eq:gbasis1}--\eqref{eq:gbasisreg}.
The theta-functions labelled by the six signs of the kinematic invariants $x_{jk}^2$ are defined in
\eqref{eq:definitiontheta} with the sign block $R$ given in \eqref{eq:signblock}.
Alternatively, the result \eqref{eq:Boxans} can be expressed in terms of the regularised functions $g^{+}_j$, cf.\ \eqref{eq:altans}.
We would like to stress that this novel compact formula for the box integral is valid in all 64 kinematic regions of Minkowski space, as verified by explicit comparison with the one-loop evaluation package \cite{vanOldenborgh:1990yc}.

\section{Conclusions and Outlook}
\label{sec:outlook}

In the present paper we considered an extension of the approach to bootstrap Feynman integrals using integrability from Euclidean to Minkowski space.
Employing only its symmetries, we have constrained the Minkowski box integral in all 64 kinematic regions up to six constant coefficient vectors that contain 
twelve free parameters. The latter can be fixed using analytic continuation as discussed in \Secref{sec:FixConstants}, or alternatively, by computing the integral for a few fixed configurations of the external kinematics, see \Appref{app:Fixingv2}. This leads to the compact result \eqref{eq:Boxans}.

The fact that some of the degrees of freedom in Minkowski space are not yet  fixed using integrability still leaves us unsatisfied. In \cite{Loebbert:2019vcj} the Euclidean integral was completely constrained via its symmetries, together with the star-triangle relation well-known from integrability in a coincident-point limit. Using the latter, however, required to work with generic propagator powers to regularise the limit. This motivates to generalise the Minkowski space analysis of the present paper to the situation of generic propagator powers, in order to fill in this last missing piece of information.

The box integral can be considered the simplest example of the four-point ladder \cite{Usyukina:1993ch} or Basso--Dixon correlators \cite{Basso:2017jwq}. It is a natural question of whether a similarly compact formula as the Minkowski result \eqref{eq:Boxans} for the box can be given for these integrals when studied in Minkowski space.
That this is indeed the case is suggested by the universality of the arguments given in \Secref{sec:FixConstants} in the context of four-point integrals.

Another natural generalisation is to study massive Feynman integrals in Minkowski space from symmetry bootstrap, cf.\ \cite{Loebbert:2020hxk,Loebbert:2020glj}. Here the presence of masses implies a  better control over the complexity of different integrals and compact polylogarithmic formulas exist for the simplest Euclidean examples at one loop, see e.g.\ \cite{Bourjaily:2019exo}.

The box integral \eqref{eq:BoxDefinition} represents a one-loop contribution to a time-ordered correlation function in Minkowski space. It would be natural to study more general Wightman functions \cite{mack2009dindependent}, and derive formulas similar to \eqref{eq:Boxans} for arbitrary time-orderings. This would also allow for the study of the breaking of global conformal invariance in Minkowski space, as a function of time-ordering.

Throughout this paper we considered, for simplicity, the box integral for arbitrary real kinematics. The map between configurations of four points in Minkowski space and physical kinematics appears to be very complicated. In particular various `missing' and `rare' kinematic regions appear, discussed in \cite{Corcoran:2020akn} and briefly in \Appref{appendix:details}. However, we do notice that the box integral has a singular behaviour as $z\rightarrow \bar{z}$ in these regions. We wonder if this could be a criterion to classify such regions, here and at higher points.

Finally, it would be desirable to come back to our starting point in the introduction and to connect the Yangian bootstrap of Feynman integrals to computations in $\superN=4$ SYM theory. Here we note that the box integral can be understood as a correlator in the bi-scalar fishnet theory that was introduced in \cite{Gurdogan:2015csr} as a particular double scaling limit of gamma-deformed $\superN=4$ SYM theory. It would be fascinating to `reverse' these double-scaling limits and to see how the Yangian bootstrap generalizes to intermediate steps.

\subsection*{Acknowledgements}
This project has received funding from the  European Union’s Horizon 2020 research and innovation programme under the Marie Sklodowska-Curie grant agreement  No.\ 764850 “SAGEX”. The work of FL is funded by the Deutsche Forschungsgemeinschaft (DFG, German Research Foundation)--Projektnummer 363895012. JM is  supported  by  the  International  Max  Planck  Research  School  for  Mathematical  and Physical Aspects of Gravitation, Cosmology and Quantum Field Theory.

\appendix

\section{Some Details}
\label{appendix:details}
In this appendix we collect some details on the above calculations.

\paragraph{Relations Between Different Regularisations.}
In the main text, we regularised the functions $g_i$ that span the space of Yangian invariants by adding infinitesimal imaginary shifts to their arguments. The purpose of these shifts is to clarify on which side of a branch cut the function is supposed to be evaluated. Of course, different choices of the sign of the shifts do not change the value of the basis element but merely its functional representation. We therefore summarise the relations between the basis elements in different regularisations as
\begin{align}
\begin{pmatrix}g_1^{+}(z,\bar z)\\
g_2^{+}(z,\bar z)\\
g_3^{+}(z,\bar z)\\
g_4^{+}(z,\bar z)\end{pmatrix} = 
\begin{pmatrix}1 & \theta_1\bar\eta_0-\bar\theta_1\eta_0 & \theta_0\bar\eta_1 - \bar\theta_0\eta_1 & \bar\theta_1\theta_0 - \theta_1\bar\theta_0 \\
0 &1-\theta_0 - \bar \theta_0 & \theta_0 + \bar \theta_0 & - \theta_0 - \bar \theta_0\\
0 & \theta_1 + \bar \theta_1 & 1- \theta_1 - \bar \theta_1 & \theta_1 + \bar \theta_1\\
0&-\eta_0 \eta_1 - \bar \eta_0 \bar \eta_1 & \eta_0 \eta_1 + \bar \eta_0 \bar \eta_1 & -1 + \theta_0 + \bar \theta_0 + \theta_1 + \bar \theta_1
\end{pmatrix}
\begin{pmatrix}g_1^{-}(z,\bar z)\\
g_2^{-}(z,\bar z)\\
g_3^{-}(z,\bar z)\\
g_4^{-}(z,\bar z)\end{pmatrix},
\label{eq:transition}
\end{align}
where for compactness we use the notation
\begin{align}
\eta_a &= 1-\theta_a, & \theta_0 &= \theta(-z),& \theta_1 & = \theta(z-1),\\
\bar \eta_a &= 1- \bar\theta_a,& \bar \theta_0 &= \theta(-\bar z), & \bar \theta_1 &= \theta(\bar z -1).
\end{align}
The relation \eqref{eq:transition} is for $z,\bar{z}\in\mathbb{R}\setminus\{0,1\}$; of course $g_i^+$ and $g_i^-$ are indistinguishable when $z\in \mathbb{C}\setminus\mathbb{R}, \bar{z}=z^*$. We also note the relation between the function basis $f_i$ and $g_i$
\begin{align}
g_4^{\pm}= f_3^{\pm} - f_2^{\pm}\pm 2\pi if_1,
\end{align}
which is valid for all $z,\bar{z}\in\mathbb{R}\setminus\{0,1\}$. We also list some functional relations satisfied by the functions $g_{i}^{+}$ and $g_i^{-}$, for all possible $z,\bar{z}$. Here $g_1$ transforms very nicely (for the Bloch--Wigner dilogarithm this sequence of identities is also known as the 6-fold symmetry \cite{Zagier:2007knq}):
\begin{align}
    g_1^{\pm}(z,\bar{z})
    =&-g_1^{\mp}(1-z,1-\bar{z})=-g_1^{\mp}(\sfrac{1}{z},\sfrac{1}{\bar{z}})
    \label{eq:g1perm}
\nonumber\\
    =&g_1^{\pm}\left(\sfrac{1}{1-z},\sfrac{1}{1-\bar{z}}\right)=g_1^{\pm}(1-\sfrac{1}{z},1-\sfrac{1}{\bar{z}})=-g_1^{\mp}\left(\sfrac{z}{z-1},\sfrac{\bar{z}}{\bar{z}-1}\right).
\end{align}
On the other hand $g_2,g_3,$ and $g_4$ have a reduced symmetry under permutations:
\begin{align}
    g_2^{\pm}(z,\bar{z})&=-g_2^{\mp}(\sfrac{1}{z},\sfrac{1}{\bar{z}}),
     \label{eq:g2perm}
\\
    g_3^{\pm}(z,\bar{z})&=-g_3^{\mp}\left(\sfrac{z}{z-1},\sfrac{\bar{z}}{\bar{z}-1}\right),
       \label{eq:g3perm}
\\
    g_4^{\pm}(z,\bar{z})&=-g_4^{\mp}(1-z,1-\bar{z}).
       \label{eq:g4perm}
\end{align}
In general $g_{i}^\pm$ are mapped into each other under permutations for $i=1,2,3$. For example
\begin{equation}
   g_3^{\pm}(1-z,1-\bar{z})=g_2^{\mp}(z,\bar{z}).
   \label{eq:g2g3}
\end{equation}

\paragraph{Permutations.}
Under a permutation $\sigma$ of the external points, the conformal invariants transform as $z\rightarrow h_\sigma(z),\bar{z}\rightarrow h_{\sigma}(\bar{z})$ and $\phi\rightarrow \sign(\sigma)\sp\phi$. Permutations also transform the kinematic region $R$, expressed as a sign block \eqref{eq:signblock}. Modulo shuffling, permutations permute the rows of the sign block: $R\rightarrow R_{\sigma}$. For example, $\sigma=(12)$ swaps rows 2 and 3 of the sign block:
\begin{equation}
\sign \signblocktextbiggerbr{x_{12}^2,x_{34}^2}{x_{23}^2,x_{14}^2}{x_{13}^2, x_{24}^2}\rightarrow\sign  \signblocktextbiggerbr{x_{12}^2, x_{34}^2}{x_{13}^2, x_{24}^2}{x_{23}^2, x_{14}^2}.
\end{equation}
These facts are summarised in \Tabref{table:permutations}.
\begin{table}[h!]
\begin{center}
 \begin{tabular}{|c|c|c|c|} 
 \hline
 $\sigma$ &  $h_\sigma(z)$  & $R_\sigma$ \\ [0.5ex] 
 \hline

 $(\sp\sp),(12)(34),(13)(24),(14)(23)$ &  $z$& $R$\\

 \hline
$(24),(13),(1234),(1432)$ &$1-z$& $\text{row 1}\leftrightarrow\text{row 2}$ \\

 \hline
 $(23),(14),(1243),(1342)$ &  $1/z$& $\text{row 1}\leftrightarrow\text{row 3}$ \\

 \hline
 $(12),(34),(1423),(1324)$ & $ \frac{z}{z-1}$  & $\text{row 2}\leftrightarrow\text{row 3}$\\

 \hline
 $(234),(124),(132),(143)$ &  $ \frac{1}{1-z}$& $\text{row 1,2,3}\rightarrow\text{row 2,3,1}$ \\

 \hline
 $(243),(123),(134),(142)$ & $1-1/z$& $\text{row 1,2,3}\rightarrow\text{row 3,1,2}$ \\

 \hline
\end{tabular}
\caption{Transformation of box integral and conformal invariants under permutations.}
\label{table:permutations}
\end{center}
\end{table}\noindent 
Note that while permutations on the external points is an $S_4$ action, this reduces to an $S_3=S_4/(\mathbb{Z}_2\times \mathbb{Z}_2)$ action on the quantities of interest.

\paragraph{Restricting ${\phi}$ for $R_3$.}
Consider $R_3=\signblockbr{-+}{++}{++}$.
Under the permutation $\sigma=(12)$ we have $z\rightarrow z'=\frac{\bar{z}}{\bar{z}-1}$ and $\bar{z}\rightarrow \bar{z}'= \frac{z}{z-1}$. Since $\sigma R_3=R_3$ we have the constraint on the expansion \eqref{eq:Yangsatz}:
\begin{equation}
  \alpha_i^{R_3} g_{i}^{-}(z,\bar{z})+\beta_i^{R_3} g_{i}^{+}(z,\bar{z})=-\alpha_i^{R_3} g_{i}^{+}\left(\frac{z}{z-1},\frac{\bar{z}}{\bar{z}-1}\right)-\beta_i^{R_3} g_{i}^{-}\left(\frac{z}{z-1},\frac{\bar{z}}{\bar{z}-1}\right).
\end{equation}
Because of \eqref{eq:g1perm} and \eqref{eq:g3perm} this gives no constraint on $\alpha_1,\beta_1,\alpha_3,\beta_3$, however we do have $\alpha_2^{R_3}=\beta_2^{R_3}=\alpha_4^{R_3}=\beta_4^{R_3}=0$. Therefore we have
\begin{align}\label{14f}
    {\phi}(R_3)=\alpha_1^{R_3}g_{1}^{-}+\alpha_3^{R_3}g^{-}_3+\beta_1^{R_3}g_{1}^{+}+\beta_3^{R_3}g^{+}_3.
\end{align}
For the kinematic region $R_3$ we have $z\in (0,1), \bar{z}\in (-\infty,0)$, and
\begin{align}
    g_1^{-}=g_1^{+}+2\pi i g_3^{+},\spac  g_3^{-}=g_3^{+}.
\end{align}
Therefore we can write \eqref{14f} in terms of either $g_i^-$ or $g_i^+$, see \eqref{eq:A3phi} where we take $a_3= \alpha_1^{R_3}+\beta_1^{R_3}$ and $2\pi ic_3= \alpha_3^{R_3}+\beta_3^{R_3}-2\pi i \beta_1^{R_3}$.
The rest of the sector $\Lambda_3$ can be reached from $R_3$ via permutations and conjugation. For example let $R_3'=\signblockbr{++}{-+}{++}$ and $\tau=(13)$, so that $R_3'=\tau R_3$. $\tau$ generates the transformation $z\rightarrow 1-\bar{z}$ and $\bar{z}\rightarrow 1-z$. $\hat\phi$ is invariant under $\tau$, and so
\begin{align}\label{17f}
{\phi}(R_3')=-a_3 g_{1}^{+}(1-z,1-\bar{z})-2\pi ic_3g^{+}_3(1-z,1-\bar{z})=a_3 g_{1}^{-}(z,\bar{z})-2\pi ic_3g^{-}_2(z,\bar{z}),
\end{align}
and similarly
\begin{align}\label{18f}
{\phi}(R_3')=-a_3 g_{1}^{-}(1-z,1-\bar{z})-2\pi i\hat{c}_3g^{-}_3(1-z,1-\bar{z})=a_3 g_{1}^{+}(z,\bar{z})-2\pi i\hat{c}_3g^{+}_2(z,\bar{z}),
\end{align}
where we used \eqref{eq:g2g3} and that $g^{+}_2=g^-_2$ for $z\in(1,\infty), \bar{z}\in (0,1)$. These and similar equations are fed into the master formulas \eqref{eq:mastermp} and \eqref{eq:altans}.
\paragraph{Alternative Form of the Yangsatz.}

Here we specify the ansatz based on Yangian invariance as given in terms of the alternative regularised functions $g_j^+$:
\begin{alignat}{2}
\phi(z,\bar{z}|R)
=
&+a_R g_1^{+}
\nonumber\\
&+2\pi i g_2^{+}\big(&&-\hat{c}_3\theta\signblock{++}{-+}{++}+c_3^*\theta\signblock{--}{-+}{--}-\hat{c}_4\theta\signblock{-+}{--}{-+}+c_4^*\theta\signblock{-+}{++}{-+}-c_5\theta\signblock{++}{--}{++}+c_5\theta\signblock{--}{++}{--}
\nonumber\\
&&&-d_6\theta\signblock{-+}{++}{--}-b_6\theta\signblock{-+}{--}{++}-\bar{c}_6\theta\signblock{++}{-+}{--}+b_6\theta\signblock{--}{++}{-+}+d_6\theta\signblock{++}{--}{-+}-\bar{c}_6\theta\signblock{--}{-+}{++}\big)
\nonumber\\
&+2\pi i g_3^{+}\big(&&+\hat{c}_3\theta\signblock{-+}{++}{++}-c_3^*\theta\signblock{-+}{--}{--}+\hat{c}_4\theta\signblock{--}{-+}{-+}-c_4^*\theta\signblock{++}{-+}{-+}+c_5\theta\signblock{--}{++}{++}-c_5\theta\signblock{++}{--}{--}
\nonumber\\
&&&+\bar{c}_6\theta\signblock{-+}{++}{--}+\bar{c}_6\theta\signblock{-+}{--}{++}-d_6\theta\signblock{++}{-+}{--}-d_6\theta\signblock{--}{++}{-+}-b_6\theta\signblock{++}{--}{-+}+b_6\theta\signblock{--}{-+}{++}\big)
\nonumber\\
&+2\pi i g_4^{+}\big(&&-c_3\theta\signblock{++}{++}{-+}+\hat{c}_3^*\theta\signblock{-+}{--}{--}-c_4\theta\signblock{-+}{-+}{--}+\hat{c}_4^*\theta\signblock{-+}{-+}{++}-c_5\theta\signblock{++}{++}{--}+c_5\theta\signblock{--}{--}{++}
\nonumber\\
&&&
-c_6\theta\signblock{-+}{++}{--}-d_6\theta\signblock{-+}{--}{++}-b_6\theta\signblock{++}{-+}{--}-c_6\theta\signblock{--}{++}{-+}-c_6\theta\signblock{++}{--}{-+}-d_6\theta\signblock{--}{-+}{++}\big)
\nonumber\\
&&&+\theta\text{ functions},
\label{eq:altans}
\end{alignat}
where `$\theta\text{ functions}$' refers to some extra terms appearing in the $g_i^+$ expansion, for example in the second line of \eqref{eq:phiR5}. These theta functions are nonzero only in the `missing' and `rare' kinematic regions, discussed in the next section.
\paragraph{Range of $z$ and $\bar z$.}

Allowing $x_{ij}^2$ to be arbitrary real numbers, the range of $z$ and $\bar{z}$ in each kinematic region is summarised in \Tabref{table:zzbrange}. Note that for physical kinematics the table is more restricted. For example for $R=\signblockbr{++}{++}{++},\signblockbr{--}{--}{++},\signblockbr{--}{++}{--}$ or $\signblockbr{++}{--}{--}$ we cannot have $z,\bar{z}\in\mathbb{C}\setminus\mathbb{R}$. Moreover it is numerically observed that for physical configurations with $z,\bar{z}\in (1,\infty)$ it is not possible to have kinematic region $-R_5=\signblockbr{++}{--}{--}$, and that the kinematic region $R_5$ is very rare, see \cite{Corcoran:2020akn}. Such `missing' and `rare' kinematic regions are typical in the equivalence class $\Lambda_5$. In such regions the result for the box integral \eqref{eq:Boxans} blows up for $z\rightarrow \bar{z}$.

\begin{table}[h]
\renewcommand{\arraystretch}{1.2}
\begin{center}
 \begin{tabular}{|c|c|} 
 \hline
$z,\bar{z}$ Range & Kinematic Regions (Modulo Shuffling) \\ [0.5ex] 
 \hline
$\bar{z}=z^*$ or $z,\bar{z}$ same interval & $\signblockbr{--}{--}{--},\signblockbr{++}{++}{++},\signblockbr{-+}{-+}{-+},\signblockbr{+-}{+-}{+-},\signblockbr{--}{--}{++},\signblockbr{++}{++}{--},\signblockbr{--}{++}{--},\signblockbr{++}{--}{++},\signblockbr{++}{--}{--},\signblockbr{--}{++}{++}$\\
\hline
$z\in(0,1),\bar{z}\in(-\infty,0)$& $\signblockbr{-+}{++}{++},\signblockbr{-+}{--}{--},\signblockbr{++}{-+}{-+},\signblockbr{--}{-+}{-+},\signblockbr{-+}{--}{++},\signblockbr{-+}{++}{--}$\\
\hline
$z\in(1,\infty),\bar{z}\in (0,1)$&$\signblockbr{++}{-+}{++},\signblockbr{--}{-+}{--},\signblockbr{-+}{--}{-+},\signblockbr{-+}{++}{-+},\signblockbr{++}{-+}{--},\signblockbr{--}{-+}{++}$\\
\hline
$z\in(1,\infty),\bar{z}\in(-\infty,0)$&$\signblockbr{++}{++}{-+},\signblockbr{--}{--}{-+},\signblockbr{-+}{-+}{++},\signblockbr{++}{++}{-+},\signblockbr{--}{++}{-+},\signblockbr{++}{--}{-+}$\\
\hline
\end{tabular}
\caption{Dependence of $z,\bar{z}$ on kinematic region for arbitrary real kinematics. $z,\bar{z}$ `same interval' means that both $z$ and $\bar{z}$ are in $(-\infty,0), (0,1),$ or $(1,\infty)$.}
\label{table:zzbrange}
\end{center}
\end{table}
\paragraph{Equivalence Classes.}
Explicitly, the equivalence classes $\Lambda_i$ are given by
\begin{align}
&\Lambda_1=\left\{\signblockbiggerbr{--}{--}{--},\signblockbiggerbr{++}{++}{++}\right\},
\nonumber\\
&\Lambda_2=\left\{\signblockbiggerbr{-+}{-+}{-+},\signblockbiggerbr{+-}{+-}{+-}\right\},
\nonumber\\
&\Lambda_3=\left\{\signblockbiggerbr{-+}{++}{++},\signblockbiggerbr{++}{-+}{++},\signblockbiggerbr{++}{++}{-+},\signblockbiggerbr{-+}{--}{--},\signblockbiggerbr{--}{-+}{--},\signblockbiggerbr{--}{--}{-+}\right\},
\nonumber\\
&\Lambda_4=\left\{\signblockbiggerbr{--}{-+}{-+},\signblockbiggerbr{-+}{--}{-+},\signblockbiggerbr{-+}{-+}{--},\signblockbiggerbr{++}{-+}{-+},\signblockbiggerbr{-+}{++}{-+},\signblockbiggerbr{-+}{-+}{++}\right\},
\nonumber\\
&\Lambda_5=\left\{\signblockbiggerbr{--}{++}{++},\signblockbiggerbr{++}{--}{++},\signblockbiggerbr{++}{++}{--},\signblockbiggerbr{++}{--}{--},\signblockbiggerbr{--}{++}{--},\signblockbiggerbr{--}{--}{++}\right\},
\nonumber\\
&\Lambda_6=\left\{\signblockbiggerbr{-+}{++}{--},\signblockbiggerbr{++}{-+}{--},\signblockbiggerbr{--}{++}{-+},\signblockbiggerbr{-+}{--}{++},\signblockbiggerbr{--}{-+}{++},\signblockbiggerbr{++}{--}{-+}\right\}.
\label{eq:equivalence}
\end{align}


\section{Fixing Constants via Double Infinity}\label{appendix:doubleinf}
\label{app:Fixingv2}
In the main text we constrained the box integral in Minkowski space, up to 12 undetermined constants. An alternative method to fix these constants is to compute the box integral explicitly for at most four configurations in the six kinematic regions $R_i$. We identify such configurations through the use of so-called `double infinity' configurations
$X^{ab}=(x_1,x_2^b,x_3,x_4^a)$ and $Y^{ab}=(y_1,y_2^b,y_3,y_4^a)$, where $a,b\in \{+,-\}$, introduced in  \cite{Corcoran:2020akn}. 

Each of the four-vectors $x_i$ and $y_i$ have at most a single nonzero spatial component $x^3$, so we denote them in coordinates $(x^0,x^3)=(x^0,0,0,x^3)$
\begin{align}
x_1&=(0,0), 
\qquad
 x_2^b=\left(\sfrac{\xi_++\eta_b}{2},
\sfrac{\xi_+-\eta_b}{2}\right),
\qquad x_3=(1,0),
\qquad  x_4^a=\left(\sfrac{\xi_-+\eta_a}{2}, \sfrac{-\xi_-+\eta_a}{2}\right),\notag\\
y_1&=(0,0), 
\qquad
 y_2^b=\left(\sfrac{\xi_+-\eta_b}{2},\sfrac{\xi_++\eta_b}{2}\right),
\qquad
 y_3=(0,1),
\qquad  y_4^a=\left(\sfrac{-\xi_-+\eta_a}{2},\sfrac{\xi_-+\eta_a}{2}\right),
\end{align}
where $\xi_+,\xi_-\in \mathbb{R}$ and we take $\eta_{\pm}\rightarrow \pm\infty$. Geometrically this corresponds to sending two of the points to null infinity.
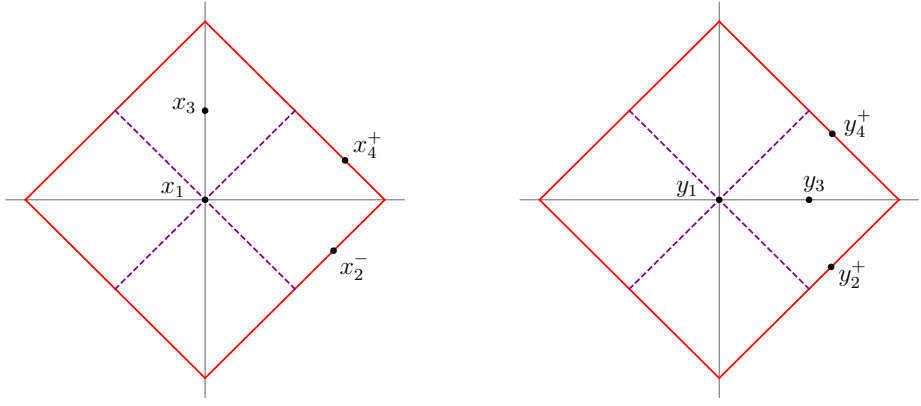
\begin{figure}[t]
    \centering
    \begin{minipage}{.5\textwidth}
        \centering
       \begin{tikzpicture}[thick,scale=0.7, every node/.style={scale=0.8}]

\node[inner sep=0pt] (russell) at (0,0)
    {\includegraphics[width=.8\textwidth]{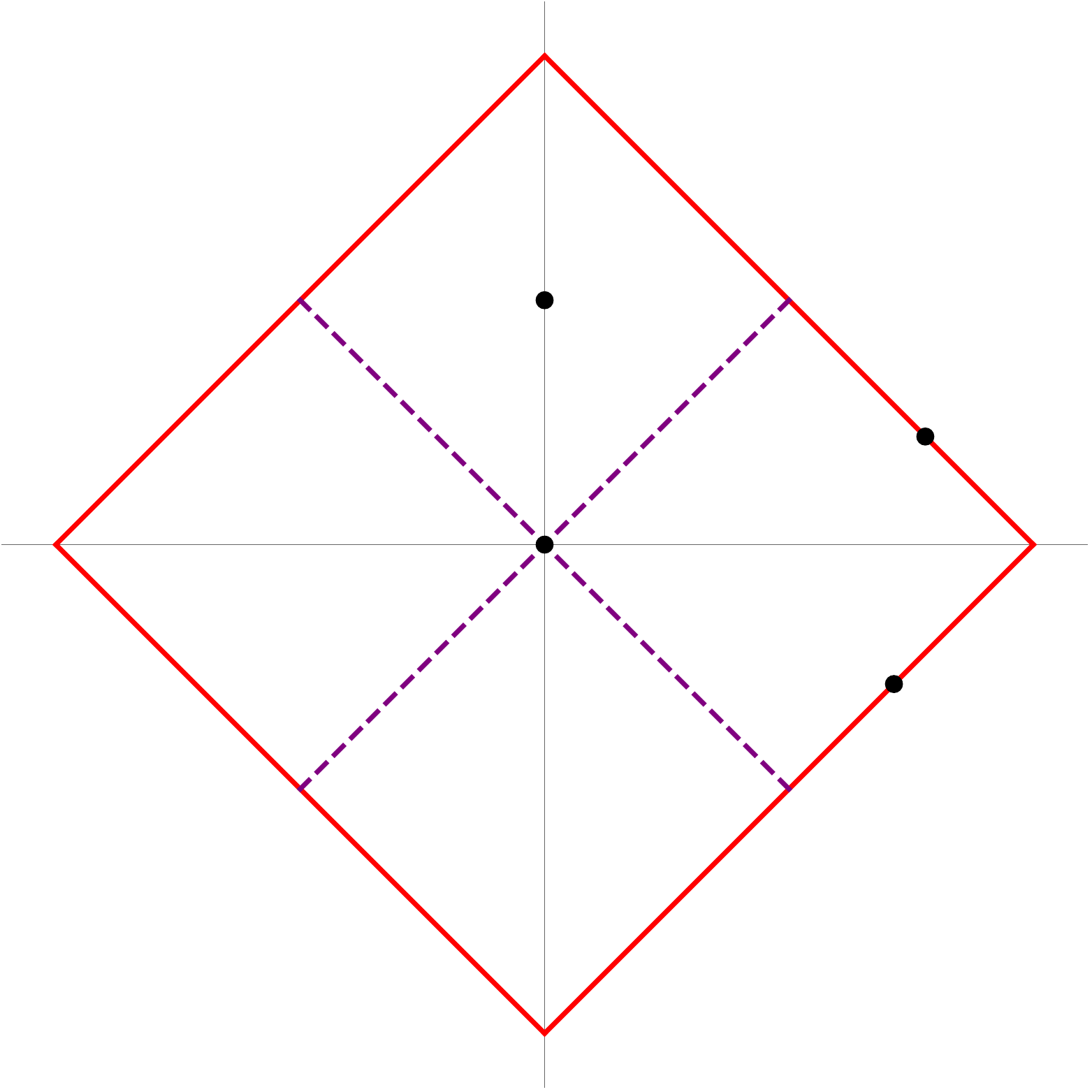}};
    \node at (-.61,0.21){$x_1$};
    \node at (2.8,-1.25){$x_2^-$};
    \node at (-.4,1.8){$x_3$};
    \node at (3.05,1.05){$x_4^+$};
\end{tikzpicture}
    \end{minipage}%
    \begin{minipage}{0.5\textwidth}
        \begin{tikzpicture}[thick,scale=0.7, every node/.style={scale=0.8}]

\node[inner sep=0pt] (russell) at (0,0)
    {\includegraphics[width=.8\textwidth]{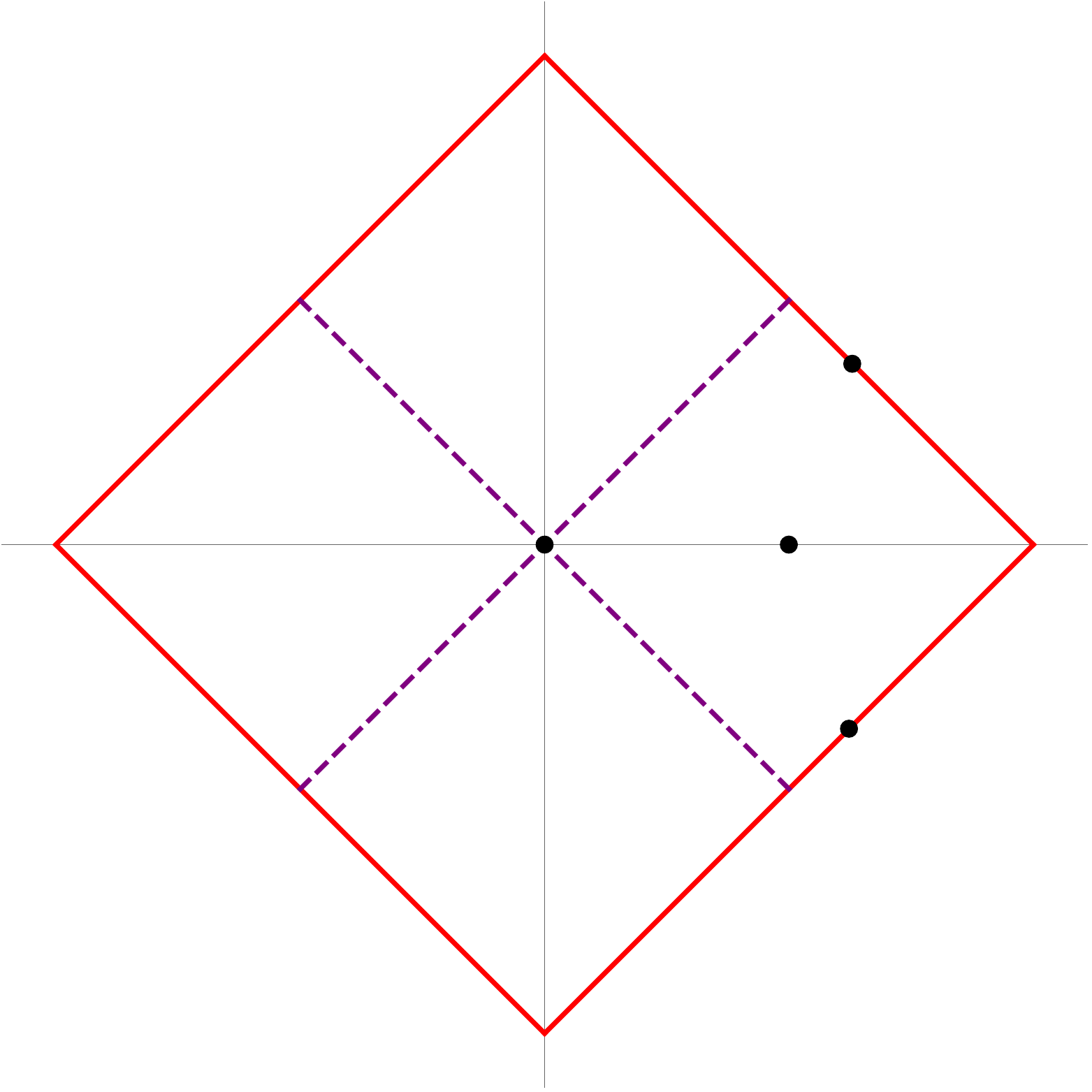}};
    \node at (-.61,0.21){$y_1$};
    \node at (2.5,-1.4){$y_2^+$};
    \node at (1.78,0.3){$y_3$};
    \node at (2.6,1.45){$y_4^+$};
\end{tikzpicture}
    \end{minipage}
    \caption{Configurations $X^{+-}$ and $Y^{++}$ on a Penrose diagram.}
    \label{figure:doubleinf}
\end{figure}
We give examples of these configurations in \Figref{figure:doubleinf}. For each of these configurations the cross-ratios are
\begin{equation}
u=\xi_+(1-\xi_-),
\qquad\qquad
 v=\xi_{-}(1-\xi_+).
\end{equation}
The kinematic region of these configurations is 
\begin{equation}
R(X^{ab})= 
 \sign \signblocktextbiggerbralt{\xi_+\eta_b}{\eta_a(\xi_--1)}{\eta_b(\xi_+-1)}{\xi_-\eta_a } {1 }{ -\eta_a\eta_b} ,
\qquad\qquad
R(Y^{ab})=-R(X^{ab}).
\label{eq:RY}
\end{equation}
By varying $a,b$ and the range of $\xi_+,\xi_-$ one can access a large number of kinematic regions with the configurations $X$ and $Y$. We show how to access the kinematic regions $R_i$ using these configurations in \Tabref{table:Rdoubleinf}.
\begin{table}[t]
\begin{center}
 \begin{tabular}{|c|c|c|c|} 
 \hline
 Kinematic Region & Configuration & $\xi_+$ range & $\xi_-$ range \\ [0.5ex] 
 \hline
$R_1$& $Y^{+-}$ & $(-\infty,0)$ & $(1,\infty)$\\
\hline
$R_2$&$Y^{++}$ &$(1,\infty)$ & $(-\infty,0)$\\
\hline
$R_3$&$X^{+-}$ &$(0,1)$ &$(1,\infty)$\\
\hline
$R_4$&$Y^{--}$ &$(-\infty,0)$ &$ (0,1)$\\
\hline
$R_5$&$X^{+-}$ &$(0,1)$ & $(0,1)$\\
\hline
$R_6$&$Y^{-+}$ &$(0,1)$ &$ (1,\infty)$\\ 
\hline

\end{tabular}
\caption{How to realise kinematic regions $R_i$ with double infinity configurations.}
\label{table:Rdoubleinf}
\end{center}
\end{table}
It is possible to compute the box integral for the configurations $X^{ab}$ and $Y^{ab}$ directly in Minkowski space. We define
\begin{align}
I^{ab}_{X}(\xi_+,\xi_-)= I_4(X^{ab}),\spac  I^{ab}_{Y}(\xi_+,\xi_-)= I_4(Y^{ab}).
\end{align}
Using \eqref{eq:conjugation} and \eqref{eq:RY} it is clear that
\begin{equation}
I^{ab}_{Y}(\xi_+,\xi_-)=I^{ab}_{X}(\xi_+,\xi_-)^*.
\label{eq:doubleinfconj}
\end{equation}
Moreover, using permutation covariance and translation, parity, and time-reversal invariance one can show \begin{equation}
    I^{-+}_{X}(\xi_+,\xi_-)=I^{+-}_{X}(\xi_-,\xi_+),\spac  I^{++}_{X}(\xi_+,\xi_-)=I^{--}_{X}(1-\xi_+,1-\xi_-). 
    \label{eq:doubleinfsym}
\end{equation}
Therefore, calculating the box integral for the 8 configurations $X^{ab}$ and $Y^{ab}$ is reduced to just calculating it for $X^{+-}$ and $X^{--}$. The box integral in the remaining configurations can then be recovered using \eqref{eq:doubleinfconj} and \eqref{eq:doubleinfsym}. The box integral \eqref{eq:BoxDefinition} in these configurations simplifies considerably. For example
\begin{equation}
    I^{+-}_{X}= \frac{2i}{\pi}\int_{x,r,t} \frac{r^2}{(t^2-r^2+i\epsilon)((t-1)^2-r^2+i\epsilon)(\xi_+-t-rx-i\epsilon')(\xi_--t+rx+i\epsilon')},
    \label{equation:boxdoubleinf}
\end{equation}
where $\epsilon'\ll\epsilon$ and $\int_{x,r,t}=\int_{-1}^1 dx\int_{0}^{\infty}dr\int_{-\infty}^\infty dt $. This computation was done in \cite{Corcoran:2020akn}, via a simple calculation of residues. After completing this calculation (and a similar one for $I^{--}_X$), we can evaluate the box integral in the kinematic regions $R_i$ (see  \Tabref{table:Rdoubleinf}) for sufficiently many points to fix the free constants in \eqref{eq:mastermp}. 

\pdfbookmark[1]{\refname}{references}
\bibliographystyle{nb}
\bibliography{MinkowskiBox}


\end{document}